\documentclass[a4paper]{report}
\usepackage[utf8]{inputenc}
\usepackage[T1]{fontenc}
\usepackage{RJournal}
\usepackage{amsmath,amssymb,array}
\usepackage{booktabs}
\usepackage{float}

\begin{document}

\sectionhead{Contributed research article}
\volume{XX}
\volnumber{YY}
\year{20ZZ}
\month{AAAA}

\begin{article}
\title{reslr: An R package for relative sea level modelling}

\author{by Maeve Upton, Andrew Parnell, and Niamh Cahill}

\maketitle

\abstract{%
We present \texttt{reslr}, an R package to perform Bayesian modelling of relative sea level data. We include a variety of different statistical models previously proposed in the literature, with a unifying framework for loading data, fitting models, and summarising the results. Relative sea-level data often contain measurement error in multiple dimensions and so our package allows for these to be included in the statistical models. When plotting the output sea level curves, the focus is often on comparing rates of change, and so our package allows for computation of the derivative of sea level curves with appropriate consideration of the uncertainty. We provide a large example dataset from the Atlantic coast of North America and show some of the results that might be obtained from our package.
}

\hypertarget{introduction}{%
\section{Introduction}\label{introduction}}

Understanding the rates and spatial patterns of Relative Sea-Level (RSL) change across various timescales, spanning from decades to millennia, poses a significant challenge. The task involves analysing sparse and noisy proxy and/or instrumental data sources that often have large measurement uncertainties. To address these complexities and provide robust assessments, statistical models play a pivotal role and have become indispensable in the task of quantifying RSL changes (as examined by \citet{Cahill2015_cp} and \citet{Khan2015}) and in the evaluation of temporal and spatial variability (e.g., \citet{Kopp2013}; \citet{Kopp2016}; \citet{Kemp2018}; \citet{Walker2021}). To that end the paleo sea-level community would benefit from a comprehensive toolset capable of analysing the historical evolution of sea-level changes across different times and locations. This motivated us to create the \texttt{reslr} package, which is available on the Comprehensive R Archive Network at \url{https://cran.r-project.org/web/packages/reslr} or on GitHub at \url{https://github.com/maeveupton/reslr}. Our package includes a suite of statistical models appropriate for modelling the complexity of sea-level over time and space, while accounting for sea-level data uncertainties and remaining computationally tractable. The output of our package provides insight into temporal and spatial sea-level variability and rates of sea-level change.

The \texttt{reslr} package includes a comprehensive dataset of proxy RSL reconstructions for 21 locations along the Atlantic coast of North America \citep{Kemp2013SealevelUSA}. These reconstructions rely heavily on dated geological archives obtained from coastal sediments \citep[e.g.][]{Gehrels1994} or corals \citep[e.g.][]{Meltzner2017}. Moreover, users of the package have the option to incorporate instrumental sea-level data sourced from the Permanent Service Mean Sea Level online database, which provides annual RSL measurements for approximately 1,500 tide-gauge stations worldwide \citep{Holgate_PSMSL2013}. By offering this diverse range of data sources, the \texttt{reslr} package caters to the needs of researchers seeking to explore and analyse sea-level variations across different locations and time periods.

The \texttt{reslr} package offers a range of statistical models which include: linear regression \citep[e.g.,][]{Ashe2019}, change point models \citep[e.g.][]{Cahill2015_cp}, integrated Gaussian process (IGP) models \citep[e.g.,][]{Cahill2015aStats}, temporal splines \citep[e.g.,][]{deBoor1978}, spatio-temporal splines \citep[e.g.,][]{simpson2018modelling} and generalised additive models (GAM) \citep[e.g.,][]{Upton2023noisy}. In all cases, a Bayesian framework is employed, facilitating the estimation of unknown parameters based on the RSL data while fully accounting for the associated uncertainties. The \texttt{reslr} package enables researchers to gain comprehensive insights into sea-level variations, leveraging the flexibility and robustness of these statistical models.

When it comes to addressing measurement uncertainty in proxy records, the \texttt{reslr} package offers two distinct approaches. The first approach involves employing the Errors-in-Variables (EIV) method, which takes into account the inherent uncertainties in the input variables \citep{Dey2000}. This method acknowledges that the input variables are not error-free and incorporates this knowledge in the analysis. The second approach offered by the package is the Noisy Input (NI) uncertainty method. This method tackles uncertainty by inflating the output noise variance with a corrective term that is directly linked to the input noise variance \citep{McHutchon2011}. Both the EIV and NI methods have their respective advantages, and the \texttt{reslr} package recommends the most suitable uncertainty method based on the statistical model being employed. This ensures that researchers can select the appropriate approach to effectively address measurement uncertainties within their specific analysis context.

For each model, the \texttt{reslr} package generates informative plots illustrating the model-based estimates of RSL. In the case of more complex models like the IGP, splines, and GAMs, the resulting plots not only provide RSL estimates but also offer insights into the rates of RSL change. Of particular significance to the paleo-sea level community, the GAM model provides estimates for separate components that represent potential drivers of RSL change. This feature enables comparisons between different components and contributes to a more comprehensive understanding of the factors influencing RSL fluctuations \citep{Upton2023noisy}. These visual representations enable researchers to gain a clearer understanding of when and where RSL changes occurred, including the magnitude of their temporal variations. Moreover, the package grants users access to the posterior samples used to generate the plots, providing the option to delve deeper into the underlying statistical distributions and uncertainties associated with the estimated RSL changes. The combination of these outputs serves as a valuable resource for researchers, aiding in the investigation and interpretation of RSL dynamics across various spatial and temporal contexts.

Our paper has the following structure. Firstly, we introduce the example dataset provided within the package, which serves as the foundation for the examples presented throughout the paper. We also provide insight into additional data sources. Secondly, we offer an overview of the statistical models available in the package, providing necessary background information. Next, we explore the uncertainty methods employed within these statistical models. Following this, we provide a detailed description of the functionality of the \texttt{reslr} package, outlining the diverse outputs and plots accessible to users. Finally, we conclude with important remarks and discuss potential future extensions for the package's advancement. Whilst this paper is just a summary of the features of \texttt{reslr}, a more complete vignette containing examples of the full functionality of the package is available at \url{https://maeveupton.github.io/reslr/}.

\hypertarget{background}{%
\section{Data and Models}\label{background}}

\hypertarget{data-sources}{%
\subsection{Data sources}\label{data-sources}}

Proxy sea-level data are vital sources of information for examining historic changes in RSL prior to the instrumental data period. A proxy refers to a characteristic that can be observed and used to estimate a variable of interest, which cannot be measured directly, and can be of physical, biological or chemical nature \citep[e.g.,][]{Gornitz2009}. In sea-level studies, the proxy data can be sources from microorganisms such as foraminifera \citep[e.g.,][]{Edwards2015_SLhandbook}, geochemical measurements \citep[e.g.,][]{Marshall2015_SLhandbook}, or vegetation that have accumulated in the tidal realm \citep[e.g.,][]{Kemp2015_SLhandbook}. The data sets we use have had their proxy measurements transformed into sea level using various techniques which are beyond the scope of our paper \citep[e.g.,][]{Gehrels1994,Shennan2015_Handbook,Kemp2018}. In the \texttt{reslr} package, we provide an example proxy dataset which contains 21 proxy sea-level records (See Appendix) from the Atlantic coast of North America as used in \citet{Upton2023noisy}.

Within the context of sea-level analysis, instrumental data plays an important role by providing direct measurements obtained from tide gauges and satellites (although the latter are currently not incorporated into the \texttt{reslr} package). To enhance the versatility of the package, we have implemented a feature that allows users to download annual tide-gauge data from the PSMSL Level online database and store it in a temporary file, making it readily available when needed \citep[PSMSL, 2023][]{Holgate_PSMSL2013,Woodworth2003}.

To ensure the comparability of the tide-gauge data with proxy records, we apply two processing steps. Firstly, the tide-gauge data in the PSMSL database is given in millimetres relative to a revised local reference datum (a coordinate system which defines the zero level for sea level measurements \citet{pugh_woodworth_2014}). Within \texttt{reslr}, we transform the data by removing 7000 mm to revert the tide-gauge data into the observed reference frame and convert the RSL to metres following the guidance from the PSMSL website as described in \citet{PSMSLinstruction}. The second processing step involves averaging the tide-gauge data to equation with the resolution of the temporal resolution of the more recent proxy data. However, we provide flexibility for users to adjust this averaging period according to the specific characteristics of their data.

\hypertarget{statisticalmodel}{%
\subsection{Statistical Models}\label{statisticalmodel}}

Within the \texttt{reslr} package, a Bayesian hierarchical framework is employed for each statistical modelling technique. Markov Chain Monte Carlo (MCMC) simulations are carried out using the Just Another Gibbs Sampler (JAGS) tool \citep{plummer2003jags} and implemented using the \texttt{rjags} package in R \citep{plummer2016rjags}.

Mathematically, the data level for every statistical model is described as:
\begin{equation}
y = f(\mathbf{x},t) + \epsilon_y
\end{equation}
where \(y\) is the response data (RSL in metres). \(f(\mathbf{x},t)\) is the process mean that depends on location \(\mathbf{x}\) and time \(t\). \(\epsilon_y\) is the error term given by \(\epsilon_y \sim \mathbb{N}(0, \sigma_y^2 + s_y^2)\), where \(\sigma_y^2\) is the residual variance and \(s_y\) the known measurement error associated with RSL. In Table 1 we provide a list of all the possible options for \(f\) within the \texttt{reslr} package. Since some of the models we fit do not vary over space (they apply to a single site or treat a set of sites as identical) we use \(f(t)\) rather than \(f(\mathbf{x}, t)\) to denote the process model.

When using proxy RSL data, measurement error is also present in the input variable (time) due to the dating technique used. For the input measurements, \(\tilde{t}\) is assumed to be a noisy estimate of the true time value \(t\):
\begin{equation}
 \tilde{t} = t + \epsilon_t
\end{equation}
with the error term given by \(\epsilon_t \sim \mathbb{N}(0, s_t^2)\) where \(s_t\) is the known measurement error associated with time.

We use two methods to account for the time measurement uncertainty. The first is the Errors-in-variables (EIV) method which assumes that the input variable, e.g. time, is measured as an error-prone substitute and models it directly \citep{Dey2000}. The second uncertainty method is the Noisy Input method. This method fits an initial model and uses the derivative of the mean of \(f\) to calculate a corrective variance term. Then, the model is re-run with this additional corrective variance term allowing for the input noise variation to be learned from the complete outputs of the model \citep{McHutchon2011}. Within the \texttt{reslr} package, the EIV method is used for the linear regression, change point and IGP model and the temporal spline, the spatio-temporal spline and the GAM use the Noisy Input uncertainty method. In general, the EIV method tend to be slower but models the uncertain input process directly, whilst the NI method is faster but requires the model to be fitted twice.

In Table 1, we present a range of statistical modelling techniques for \(f(\mathbf{x},t)\) the component of our approach, available in the \texttt{reslr} package. Below we discuss each technique and provide insight into the potential uses of these techniques for the paleo-environmental community.

\begin{table}
\centering
\begin{tabular}{p{30mm}p{50mm}p{20mm}}
 \textbf{Statistical Model} & \textbf{Model Information} & \textbf{\texttt{model\_type} code} \\
 \hline
Errors in variables simple linear regression & A straight line of best fit taking account any age and measurement errors in the RSL values using the method of \citet{Cahill2015_cp} & \texttt{"eiv\_slr\_t"} \\
Errors in variables change point model & An extension of the linear regression modelling process. It uses piece-wise linear sections and estimates where/when trend changes occur in the data \citep{Cahill2015_cp} & \texttt{"eiv\_cp\_t"} \\
Errors in variables integrated Gaussian process & A non-linear fit that utilities a Gaussian process prior on the rate of sea-level change that is then integrated \citep{Cahill2015aStats}. & \texttt{"eiv\_igp\_t"} \\
Noisy Input spline in time & A non-linear fit using regression splines \citep{Upton2023noisy}. & \texttt{"ni\_spline\_t"} \\
Noisy Input spline in space and time & A non-linear fit for a set of sites across a region using the method of \citet{Upton2023noisy}. & \texttt{"ni\_spline\_st"} \\
Noisy Input Generalised Additive model for the decomposition of the RSL signal & A non-linear fit for a set of sites across a region and provides a decomposition of the signal into regional, local linear and non-linear local components. This full model is as described in \citet{Upton2023noisy}. & \texttt{"ni\_gam\_decomp"} \\
\hline
\end{tabular}
\caption{List of all statistical models available in the \texttt{reslr} package. We provide a short description and the relevant literature for each model. The \texttt{model\_type} code column represents the text input the user should use when implementing their preferred modelling technique.\label{Tab:stats_mod}}
\end{table}

\hypertarget{linearregression}{%
\subsection{EIV Linear Regression}\label{linearregression}}

The EIV linear regression model is given by:
\begin{equation}
f(t) = \alpha + \beta t
\end{equation}
where \(\alpha\) is the intercept, \(\beta\) is the slope and \(t\) is time. Earlier studies, for example \citet{Shennan2002} and \citet{Engelhart2009}, employed linear regression when evaluating the rate of RSL change over the past 4000 years. The \texttt{reslr} package implements a temporal linear regression as its simplicity is popular for approximate estimates of linear rates of RSL change. However, linearity assumptions for RSL change are often unrealistic when examining long-term historical trends.

\hypertarget{changepointmodel}{%
\subsection{EIV Change Point Model}\label{changepointmodel}}

The EIV change point (CP) model, an extension of the linear regression model, assumes the RSL process is piecewise linear and estimates when trend changes occur in the data \citep{Cahill2015_cp}. Mathematically, the multiple CP model, \(f(t)\) is described as:
\begin{equation}
     f(t) = 
     \begin{cases}
      \alpha_1 + \beta_j(t - \lambda_1) \text{ when } j = 1,2,\\
      \alpha_{j-1} + \beta_j(t - \lambda_{j-1}),  \text{ when } j = 3,..., m+1
    \end{cases}
 \end{equation}
where \(t\) is time. \(\lambda_l\) is the time at which the CP occurs with the prior constraint that \(\lambda_1 < ... < \lambda_m\) where \(m\) is the number of CPs. In the \texttt{reslr} package, the user can select \(m\) to be 1, 2 or 3 CPs. \(\alpha_1\) is the expected value of RSL at the CP. \(\beta_j\) is the slope before and after the CP \citep{Cahill2015_cp}.

This technique has be used in different aspects of the sea-level literature. For example, \citet{Kemp2009} determined the magnitude and the timing of recent accelerated sea-level rise using change point models in North Carolina, USA. \citet{Brain2012} used the CP method to examine the impacts of sediment compaction on reconstructing recent sea-level rise in the United Kingdom. \citet{Hogarth2020}) used CP models to obtain more consistent estimates of sea-level rise since 1958 for the British Isles. The main advantage of the CP model is its' ability to identify sudden changes in RSL. However, the number of change points must be specified by the user.

\hypertarget{integratedgaussianprocess}{%
\subsection{Integrated Gaussian Process}\label{integratedgaussianprocess}}

An Integrated Gaussian process (IGP) is a modelling strategy that has been used extensively by the sea-level community when examining the temporal evolution of sea level change \citep[e.g.,][]{Cahill2015aStats,Hawkes2016,Kemp2017,Shaw2018,Dean2019,Stearns2023,Kirby2023}.

The IGP uses Gaussian Process (GP) to directly estimate the rate of change of the response \citep{Holsclaw2013}. In order to extract the original \(f(t)\) we integrate \(p(t)\):
\begin{equation}
    f(t) = \alpha + \int^t_0p(u)du
\end{equation}
where \(\alpha\) is the intercept and is the rate of change, \(p(t) = \frac{df}{dt}\), described as:
\begin{equation}
    p(t) \sim GP(\mu(t),k(t,t'))
\end{equation}
with \(t\) time and \(\mu(t)\) the mean function and \(k(t,t')\) is the covariance function. The covariance function provides insight into the relationship between the outcome variables, i.e.~if input variable, \(t\) and \(t'\) are in close proximity, the corresponding outcomes will be more correlated and vice versa \citep{Rasmussen2006}. It is written as \citep{Cahill2015aStats}:
\begin{equation} 
    k(t,t') = \nu^2\rho^{(t-t')^2}
\end{equation}
where \(\rho\) is the correlation parameter and \(\nu^2\) is the variance of the rate process.

The technique, described by \citet{Cahill2015aStats}, offers insights into examining rates of Relative Sea Level (RSL) change using proxy records from a single location. Apart from the IGP model, the \texttt{reslr} package does not rely on GP methods. We acknowledge that the use of GP modelling has gained considerable traction within the sea-level research community, particularly for investigating the spatio-temporal evolution of sea-level changes, as evidenced by notable studies \citep[e.g.,][]{Kopp2009,Kopp2013,Kopp2016,Kemp2018,Walker2021}. Nevertheless, in the context of the \texttt{reslr} package, we have intentionally opted for computationally efficient alternatives- splines and GAMs - as detailed in our prior work \citep{Upton2023noisy}. These methods offer practical and effective approaches to analysing sea-level data, accommodating the complexities of spatio-temporal dynamics while ensuring computational tractability.

\hypertarget{temporalspline}{%
\subsection{Temporal Spline}\label{temporalspline}}

Splines are mathematical tools used in a wide range of settings from interpolation to data smoothing. There are a variety of different splines available, yet in this research we focus on B-splines \citep{deBoor1978,Dierckx1995} and P-splines \citep{eilers_1996}. Mathematically, B-splines are described in the following way:
\begin{equation}
    f(t) = \sum^K_{k=1} b_k(t) \beta_k 
\end{equation}
where \(b_k(t)\) is the spline basis function and \(\beta_k\) is the spline coefficient.

Following on from B-splines, \citet{eilers_1996} describe a method to overcome the difficulty of choosing the correct number of knots by developing penalised spline or P-splines. Penalised differences in the spline coefficients control the smoothness of the spline based on differences (of order \(d\)) of the spline coefficients. The first order differences are written as:
\begin{equation}
\Delta\beta_k = \beta_k - \beta_{k-1}
\end{equation}
The spline coefficient will be centered on the previous value with a smoothness parameter \(\sigma^2_{\beta}\):
\begin{equation}
\Delta \beta_k \sim \mathcal{N}(0, \sigma_{\beta}^2)
\end{equation}

In our package, P-splines are used for the NI spline in time and the extendable nature of these splines allows for different components to be examined within the GAM which is described below.

\hypertarget{spatialtemporalspline}{%
\subsection{Spatio-Temporal Spline}\label{spatialtemporalspline}}

We use a spatio-temporal spline to examine RSL evolving over time at multiple locations. We include a tensor product to capture the variability over time and space (represented with longitude and latitude). For each individual covariate, time (\(t\)) and longitude (\(x_1\)) and latitude (\(x_2\)), we construct a B-spline basis \citep{Wood2017pspline}. These basis functions are combined product-wise in the following way \citep{Wood2006a}:
\begin{equation}
  f(t,x_1,x_2) = \sum_{h=1}^{H}\sum_{i=1}^{I}\sum_{j=1}^{J} b_h(t) b_i(x_1) b_j(x_2) \beta_{hij}
\end{equation}
where \(\beta_{hij}\) is the spline coefficient. \(H\) is the number of knots for \(b_h(t)\) the spline basis function in time \(t\). \(I\) is the number of knots for \(b_i(x_1)\) the spline basis function for longitude. \(J\) is the number of knots for \(b_j(x_2)\) the spline basis functions for latitude values. The prior for the spline coefficient is given as:
\begin{equation}
\beta_{hij} \sim \mathcal{N} (0, \sigma^2_{\beta})
\end{equation}
where \(\sigma^2_{\beta}\) is the smoothness parameter for the spatio-temporal spline. The \texttt{reslr} package uses B-splines for the NI spline in space and time allowing for multiple sites to be examined. The advantage of the tensor B-spline approach is that the basis functions are simple to construct, each depending on only one input variable. However the number of parameters to estimate does increase considerably.

\hypertarget{generalisedadditivemodels}{%
\subsection{Generalised Additive Models}\label{generalisedadditivemodels}}

Generalised additive models are an extension of generalised linear models that use a basis expansion and a smoothing penalty to create linear predictors that are dependent on the sum of smooth functions of the predictor variable \citep[GAMs;][]{wood_2017}. The model developed by \citet{Upton2023noisy} uses splines and random effects to create a spatio-temporal relative sea level field. It identifies variations of sea-level at different spatial and temporal scales, encompassing multiple underlying processes and avoiding a focus on specific physical processes. The decomposition of this mean relative sea level field can be written as:
\begin{equation}
f(\mathbf{x},t) =  r(t)+ g(z_\mathbf{x}) + h(z_\mathbf{x}) + l(\mathbf{x},t)
    \end{equation}
where \(r(t)\) is the regional component at time \(t\) represented with a spline in time. \(g(z_{\mathbf{x}})\) is the linear local component at location \(x\) represented by a random effect with \(z_{\mathbf{x}}\) representing each data site. \(h(z_\mathbf{x})\) is the spatial vertical offset for each data site. \(l(\mathbf{x},t)\) is the non-linear local component represented with a spline in space time.

The regional component (\(r(t)\)) represents temporal processes that are common to all locations, including barystatic and thermosteric contributions, where the former is caused by the transfer of mass between land-based ice and oceans \citep{Gregory2019} and the latter is influenced by changes in global temperature creating density variations within the oceans \citep{Grinsted2015}. It is described using a spline in time:
\begin{equation}
r(t) = \sum^{k_r}_{s=1} b_{r_s}(t)\beta^{r}_s
\end{equation}
where \(\beta^{r}_s\) is the \(s^{th}\) spline coefficient, \(k_r\) is the number of knots and \(b_{r_s}(t)\) is the \(s^{th}\) spline basis function at time \(t\). The prior for the spline coefficients of the regional component \(\beta^{r}_s\) are:
\begin{equation}
  \beta^{r}_s \sim \mathcal{N} ( 0, \sigma_r^2)
\end{equation}
where the smoothness of the model fit is controlled by \(\sigma_r\) is the standard deviation of the spline coefficient.

The linear local component (\(g(z_x)\)) of the sea level model aims to capture linear trends present in the relative sea level signal. One such cause is glacial isostatic adjustment (GIA), which is a response of the Earth, the gravitational field, and the ocean to changes in the size of ice sheets \citep{Whitehouse2018}. On relatively short timescales, it is approximated to be linear through time with spatial variability along the Atlantic coastline of North America \citep{Engelhart2009}. Mathematically, the linear local component is an unstructured random effect for each site which is formulated as:
\begin{equation}
g(z_{\mathbf{x}_j}) = \beta^{g}_jt
\end{equation}
where \(\beta^{g}_j\) is a slope parameter specific for each site \(j\). The prior for the linear local component is given by:
\begin{equation}
\beta_j^g \sim \mathcal{N} (m_{g_j}, s^2_{g_j})
\end{equation}
where \(m_{g_j}\) and \(s^2_{g_j}\) are the empirically estimated rate and associated variance (refer to \citet{Upton2023noisy} for a detailed description).

The site-specific vertical offset \(h\) is a random effect used to capture vertical shifts associated with measurement variability between sites and is formulated as:
\begin{equation}
h(z_{\mathbf{x}_j}) = \beta^{h}_j
 \end{equation}
where \(\beta^{h}_j\) contains the random effect coefficients for site \(j\). The prior for the site-specific vertical offset \(\beta^{h}_j\) is given as:
\begin{equation}
\beta^{h}_j \sim \mathcal{N}(0, \sigma_h^2)
\end{equation}
where \(\sigma_h^2\) is the variance of the random intercept across all data sites.

The non-linear local component (\(l(\mathbf{x}, t)\)) captures structured and unstructured RSL variability on century timescales, including dynamic sea-level changes (atmospheric and oceanic circulation patterns \citep{Gregory2019}) and site-specific processes (e.g. sediment compaction affecting solid Earth's surface \citep{Horton2018}). It is described using a spatio-temporal spline function formed using a tensor product and is formulated as:

\begin{equation}
  l(\mathbf{x}, t) = \sum_{s=1}^{k_l} b_{l_s}(\mathbf{x},t) \beta^{l}_s
\end{equation}

where \(\beta^{l}_s\) is the \(s^{th}\) spline coefficient, \(k_l\) is the number of knots and \(b_{l_s}(\mathbf{x},t)\) is the \(s^{th}\) spline basis function at time \(t\) and location \(\mathbf{x}\). The prior for the spline coefficient \(\beta^{l}_s\) is given as:
\begin{equation}
\beta^{l}_s \sim \mathcal{N}(0,\sigma_l^2)
\end{equation}
where \(\sigma_l^2\) is the variance of the spline coefficients over space and time.

As described in \citet{Upton2023noisy}, B-splines are used for both the regional and local terms as this model structure balances both model usability and computational efficiency for examining proxy-based sea level reconstructions on a regional to local scale. B-splines also allow for easier prior elicitation of the smoothness parameters since they directly control the variability of the spline weights in the model.

\hypertarget{implementation}{%
\section{Implementation}\label{implementation}}

Within the package, we keep the number of functions to a minimum to ensure accessibility for users R experience. We run the statistical models using an MCMC algorithm and include a summary function to obtain a high level insight into the outputs. We use S3 classes to access the summary, print and plot commands. The package has functions to plot the input data and resulting model fits using \texttt{ggplot2} \citep{Wickham2016}. The user has access to all the underlying information used to create these plots allowing these visualisations to be re-created. In addition, the functions within the package are extendable allowing advanced users access to more complex outputs.

In this section, we provide insight into the example dataset and additional data sources using tide-gauge data within the \texttt{reslr} package. A discussion is provided into each function using two separate case studies; a single location and multiple locations. In the first case study, we demonstrate the Noisy Input temporal spline (\texttt{model\_type\ =\ "ni\_spline\_t"}) which is an example modelling strategy for a single location. In the second case study, we examine multiple locations using the Noisy Input GAM decomposition (\texttt{model\_type\ =\ "ni\_gam\_decomp"}).

\hypertarget{exampledataset}{%
\subsection{Example proxy dataset}\label{exampledataset}}

We include an example dataset called \texttt{NAACproxydata}. The full dataset with names of the locations and associated literature is in the Appendix. The \texttt{NAACproxydata} is a data frame with 1715 rows and 8 columns which include:

\begin{itemize}
\item
  Region: Region name
\item
  Site: Site name
\item
  Latitude: Latitude of the site
\item
  Longitude: Longitude of the site
\item
  RSL: Relative Sea level in metres
\item
  RSL\_err: 1 standard deviation error associated with relative sea level measured in metres
\item
  Age: Age in years Common Era (CE)
\item
  Age\_err: 1 standard deviation error associated with the age in years CE
\end{itemize}

\hypertarget{additionaldata}{%
\subsection{Including tide-gauge data}\label{additionaldata}}

The tide-gauge data available to users can be obtained from the \href{https://psmsl.org/data/obtaining/complete.php}{PSMSL online database} \citep{Holgate_PSMSL2013,Woodworth2003} through the \texttt{reslr} package. To ensure compatibility with the proxy records, several processing steps are performed within the package, as discussed earlier.

When incorporating tide-gauge data, users have three methods to select their preferred tide gauge(s). The first option is to provide a list of tide-gauge names from the PSMSL database, allowing users the freedom to select any tide gauge available. The second option is to automatically identify the nearest tide gauge to the proxy location that has more than 20 years of observations. This option proves particularly useful when examining proxy records and extending the temporal range to capture recent changes in RSL.

The final option enables the selection of all tide gauges within a 1-degree radius (latitude and longitude) of the proxy location, provided they have more than 20 years of observations. This option grants users access to a wide array of tide gauges within a larger geographic area. Moreover, users can combine the first option with either the second or the third option, allowing for the freedom to choose specific tide gauges while incorporating the nearest tide gauge or multiple tide gauges.

All the values mentioned in this paragraph are arguments that can be adjusted within the function, giving users flexibility in customizing their data selection process according to their specific requirements.

\hypertarget{casestudy1}{%
\subsection{Case Study for 1 location}\label{casestudy1}}

In the following sections, we use one site, Cedar Island North Carolina USA \citep{Kemp2011,Kemp2017}, from the example dataset, \texttt{NAACproxydata}:
\footnotesize

\begin{verbatim}
CedarIslandNC <- reslr::NAACproxydata %>% 
  dplyr::filter(Site == "Cedar Island")

glimpse(CedarIslandNC)
\end{verbatim}

\begin{verbatim}
#> Rows: 104
#> Columns: 8
#> $ Region    <chr> "North Carolina", "North Carolina", "North Carolina", "North~
#> $ Site      <chr> "Cedar Island", "Cedar Island", "Cedar Island", "Cedar Islan~
#> $ Latitude  <dbl> 34.971, 34.971, 34.971, 34.971, 34.971, 34.971, 34.971, 34.9~
#> $ Longitude <dbl> -76.38, -76.38, -76.38, -76.38, -76.38, -76.38, -76.38, -76.~
#> $ RSL       <dbl> -0.12, -0.14, -0.16, -0.18, -0.19, -0.21, -0.22, -0.23, -0.2~
#> $ Age       <dbl> 2005, 1996, 1988, 1979, 1974, 1963, 1957, 1951, 1941, 1937, ~
#> $ Age_err   <dbl> 2.25, 2.00, 5.00, 5.75, 5.50, 5.50, 7.00, 7.75, 7.75, 8.00, ~
#> $ RSL_err   <dbl> 0.06, 0.06, 0.06, 0.06, 0.06, 0.06, 0.06, 0.06, 0.06, 0.06, ~
\end{verbatim}

\normalsize

For a single location such as this case study, we recommend using an EIV IGP or a NI spline in time as they estimate RSL changes over time. In this example, tide-gauge data is not included but it is an option available to the user if they require. The next example will demonstrate a more complex analysis with the inclusion of tide gauges.

After selecting the data site from the example dataset, we use the \texttt{reslr\_load} function to process the data prior to running the statistical model and it has a number of different settings that the user can alter depending on the model choice. One such setting is the \texttt{prediction\_grid\_res} option. This provides the resolution at which predictions of RSL and RSL rates are made and subsequently plotted. We set the default at 50 years and if a finer grid is required, the user can alter the setting for \texttt{prediction\_grid\_res}. The \texttt{reslr\_load} function includes additional settings to include tide-gauge data and linear rates which will be discussed in the next case study. For the single site case study, we demonstrate the \texttt{reslr\_load} function:
\footnotesize

\begin{verbatim}
CedarIslandNC_input <- reslr_load(data = CedarIslandNC)
\end{verbatim}

\normalsize

The output of this function is a list of two dataframes called \texttt{data} and \texttt{data\_grid}. The \texttt{data} dataframe is the inputted data with an additional column called \texttt{data\_type\_id} which distinguishes proxy records from tide-gauge data. The \texttt{data\_grid} is a dataframe that is evenly spaced in time based on the \texttt{prediction\_grid\_res} value chosen by the user and is used to create the plots:
\footnotesize

\begin{verbatim}
glimpse(CedarIslandNC_input$data_grid)
\end{verbatim}

\begin{verbatim}
#> Rows: 57
#> Columns: 5
#> Groups: SiteName [1]
#> $ Longitude    <dbl> -76.38, -76.38, -76.38, -76.38, -76.38, -76.38, -76.38, -~
#> $ Latitude     <dbl> 34.971, 34.971, 34.971, 34.971, 34.971, 34.971, 34.971, 3~
#> $ SiteName     <fct> "Cedar Island,\n North Carolina", "Cedar Island,\n North ~
#> $ data_type_id <fct> ProxyRecord, ProxyRecord, ProxyRecord, ProxyRecord, Proxy~
#> $ Age          <dbl> -800, -750, -700, -650, -600, -550, -500, -450, -400, -35~
\end{verbatim}

\normalsize

A brief insight into the outputs of the \texttt{reslr\_input} function can be obtained using the print function which provides the number of observations and the sources of the data as shown below:
\footnotesize

\begin{verbatim}
print(CedarIslandNC_input)
\end{verbatim}

\begin{verbatim}
#> This is a valid reslr input object with 104 observations and  1 site(s).
#> There are  1  proxy site(s) and  0  tide gauge site(s).
#> The age units are; Common Era. 
#> Decadally averaged tide gauge data was not included. It is recommended for the ni_gam_decomp model 
#> The linear_rate or linear_rate_err was not included. It is required for the ni_gam_decomp model
\end{verbatim}

\normalsize

The next step is using the plot function to plot the raw data, shown in Figure 1, using the following:
\footnotesize

\begin{verbatim}
plot(CedarIslandNC_input, plot_caption = FALSE)
\end{verbatim}

\begin{figure}

{\centering \includegraphics[width=360px]{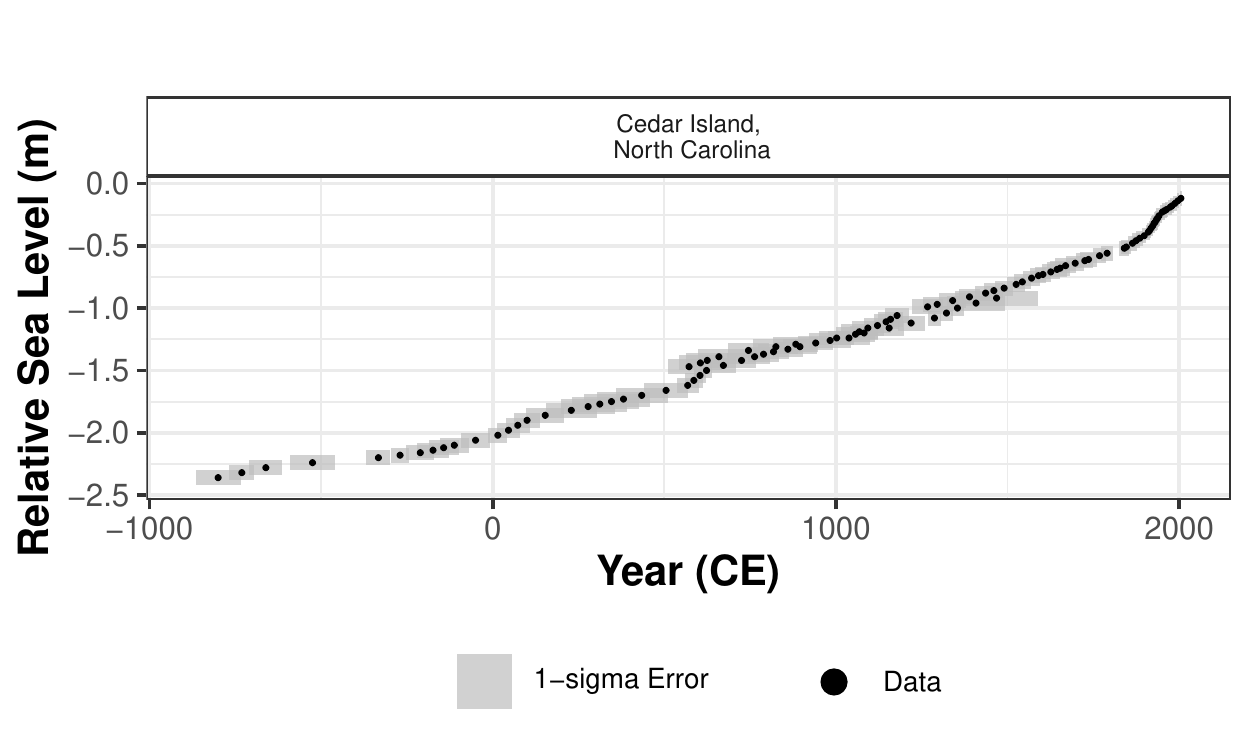} 

}

\caption{A plot of the raw data for our example site Cedar Island North Carolina \citep{Kemp2011,Kemp2017}. The x-axis is time in years in the Common Era (CE) and the y-axis is relative sea level in metres. The grey boxes are 1 standard deviation vertical and horizontal (temporal) uncertainty. The black dots are the midpoints of the uncertainty boxes.}\label{fig:plotdata}
\end{figure}
\normalsize

\hypertarget{nisplinet}{%
\subsubsection{Noisy Input Spline in time}\label{nisplinet}}

The NI spline in time (``ni\_spline\_t'') examines how the response variable, RSL, varies in time. While the EIV-IGP method is commonly used in the sea-level community, we demonstrate that the NI spline in time is a faster alternative. Unlike the Gaussian process, which has a computational complexity that grows exponentially with the number of data points, the spline equivalent uses pre-computed basis functions, resulting in a more efficient computation \citep{wood_2017}.

For this model type, we use the \texttt{reslr\_mcmc} function to implement the MCMC simulation using JAGS and the model type setting is selected to be \texttt{model\_type\ =\ "ni\_spline\_t"}.
\footnotesize

\begin{verbatim}
res_ni_spline_t <- reslr_mcmc(input_data = CedarIslandNC_input,
                              model_type = "ni_spline_t")
\end{verbatim}

\normalsize

The output of the \texttt{reslr\_mcmc} function is a list that stores the JAGS model run, the input dataframe and the dataframes for plotting the results. The user can set the size of the credible intervals by changing the \texttt{CI} setting in this function; the current default is \texttt{CI\ =\ 0.95}. In addition, the user can alter the number of iterations which will be required if the model is not converging.

To obtain a brief insight into the outputs of the \texttt{reslr\_mcmc} function, the user can use the print function which provides the number of iterations and the model type:
\footnotesize

\begin{verbatim}
print(res_ni_spline_t)
\end{verbatim}

\begin{verbatim}
#> This is a valid reslr output object with 104 observations and  1 site(s).
#> There are  1  proxy site(s) and  0  tide gauge site(s).
#> The age units are; Common Era. 
#> The model used was the Noisy Input Spline in time model.
#> The input data has been run via reslr_mcmc and has produced 3000 iterations over 3 MCMC chains.
\end{verbatim}

\normalsize

The convergence of the MCMC algorithm can be examined for the ``ni\_spline\_t'' model using the \texttt{summary} function and ensures the scale reduction factor (R-hat) is close to 1 \citep{Gelman1992,gelman2013bayesian}. If the model run has converged, the package will print: ``No convergence issues detected''. If the package prints: ``Convergence issues detected, a longer run is necessary''. The user is recommended to update the \texttt{reslr\_mcmc} function with additional iterations as described above. The \texttt{summary} function provides insight into the parameter estimates from the model using the following:
\footnotesize

\begin{verbatim}
summary(res_ni_spline_t)
\end{verbatim}

\begin{verbatim}
#> No convergence issues detected.
\end{verbatim}

\begin{verbatim}
#> # A tibble: 2 x 7
#>   variable      mean      sd     mad       q5    q95  rhat
#>   <chr>        <num>   <num>   <num>    <num>  <num> <num>
#> 1 sigma_beta 2.10    0.718   0.555   1.29     3.40    1.00
#> 2 sigma_y    0.00620 0.00482 0.00463 0.000445 0.0158  1.00
\end{verbatim}

\normalsize

For the parameter estimates, ``sigma\_beta'' acts as a smoothness parameter controlling the penalisation of the splines coefficients for the spline in time model and ``sigma\_y'' represents the data model variation. These are \(\sigma_y\) and \(\sigma_{\beta}\) as described in Section 2.

The final results from the ``ni\_spline\_t'' model can be illustrated using the \texttt{plot} function and the corresponding dataframes are stored in the \texttt{res\_ni\_spline\_t} object called \texttt{output\_dataframes} as a named list element. Figure 2 demonstrates the posterior model fit for our example site using:

\footnotesize

\begin{verbatim}
plot(res_ni_spline_t,
     plot_type = "model_fit_plot",
     plot_caption = FALSE)
\end{verbatim}

\begin{figure}[H]

{\centering \includegraphics[width=432px]{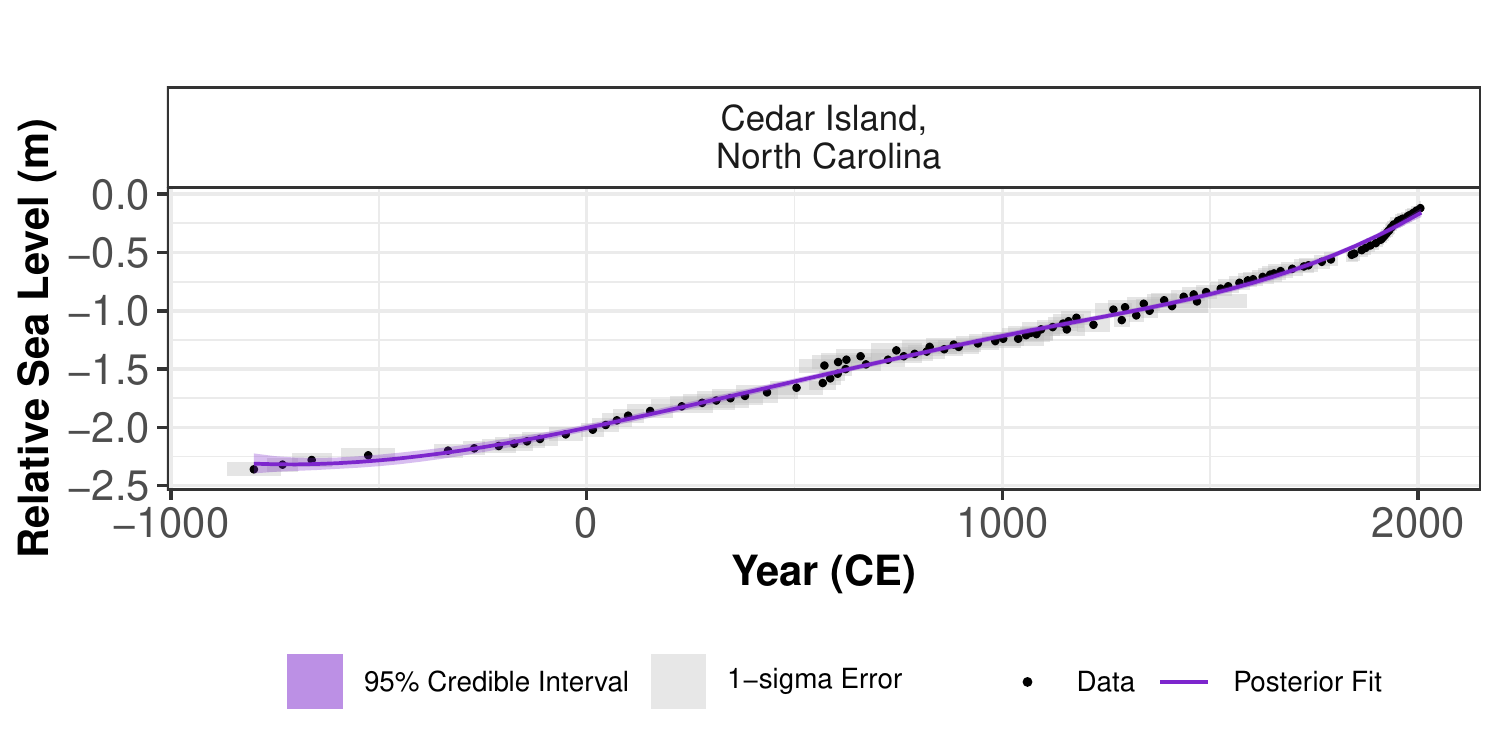} 

}

\caption{The plot of the noisy input spline in time model fit for our example site, Cedar Island, North Carolina \citep{Kemp2011,Kemp2017}. The x-axis is time in years in the Common Era (CE) and the y-axis is relative sea level in metres. The grey boxes are 1 standard deviation vertical and horizontal (temporal) uncertainty. The black dots are the midpoints of the uncertainty boxes. The solid purple line represents the posterior model fit with a 95\% credible interval denoted by shading.}\label{fig:nisplinemodfit}
\end{figure}

\normalsize

In Figure 3, the rate of change of this posterior model fit is presented and can be viewed using:

\footnotesize

\begin{verbatim}
plot(res_ni_spline_t,
     plot_type = "rate_plot",
     plot_caption = FALSE)
\end{verbatim}

\begin{figure}[H]

{\centering \includegraphics[width=360px]{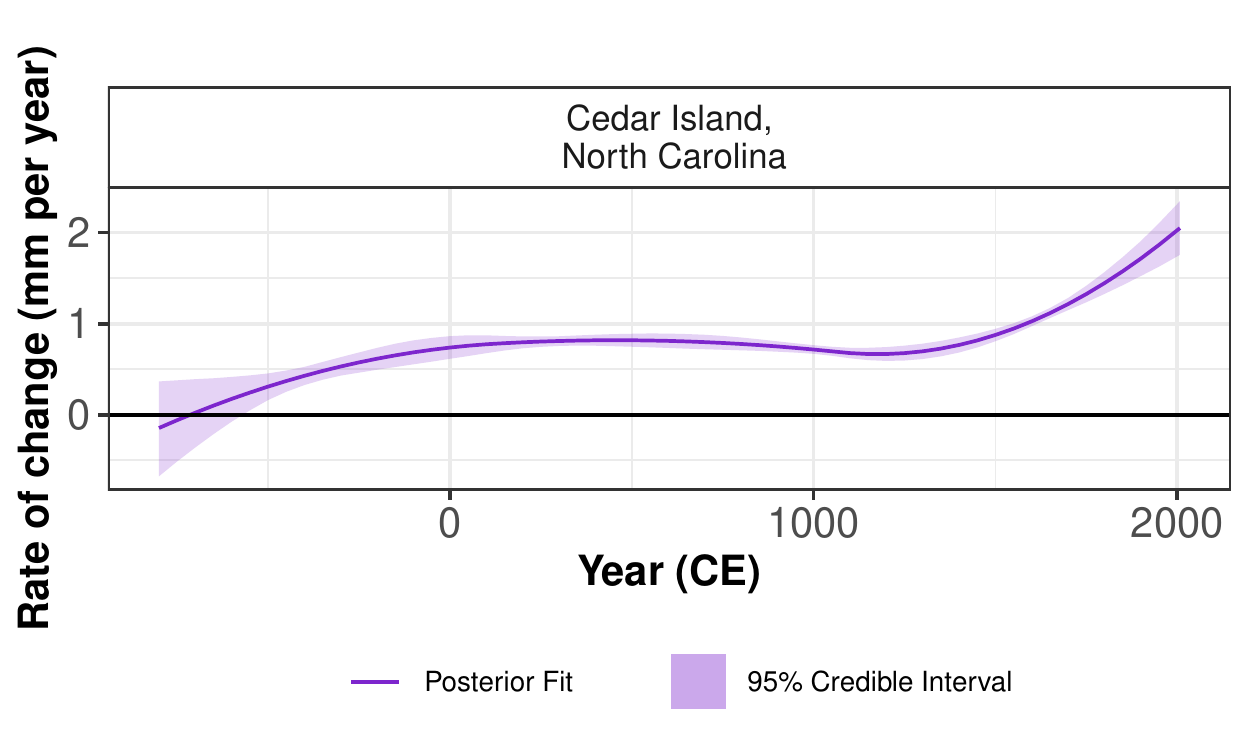} 

}

\caption{The rate of change model fit using noisy input spline in time model for our example site Cedar Island, North Carolina \citep{Kemp2011,Kemp2017}. The rate is calculated by taking the first derivative of the total model fit. The x-axis is time in years in the Common Era (CE) and the y-axis is the instantaneous rate of change of sea level in mm per year. The solid purple line represents the posterior model fit with a 95\% credible interval denoted by shading. There is a black horizontal line which is the zero rate of change for this site.}\label{fig:nisplinemodrate}
\end{figure}

\normalsize

\hypertarget{casestudy2}{%
\subsection{Case Study for multiple sites}\label{casestudy2}}

The sea-level research community is commonly interested in temporal and spatial variations in RSL. To cater to this interest, the \texttt{reslr} package offers two models for spatio-temporal modeling. The first model is a noisy-input spline that accounts for noise in both time and space, providing a robust representation of RSL dynamics. The second model, a more intricate option, is the noisy input GAM. Mathematical details concerning the Noisy Input GAM can be found in \citet{Upton2023noisy}. In our upcoming example, we will focus on this model as it empowers users to explore the decomposition of the RSL signal over time and space, unraveling valuable insights into the underlying dynamics.

\hypertarget{casestudy2nigam}{%
\subsubsection{Noisy Input Generalised Additive Model for decomposition of response signal}\label{casestudy2nigam}}

We demonstrate the functions settings required for the NI GAM. This model requires an adequate number of sites to perform the decomposition and the minimum sites required will depend on the signal in the data. In this example, we use nine sites from the example dataset, \texttt{NAACproxydata}, which are selected in the following manner:
\footnotesize

\begin{verbatim}
multi_site <- reslr::NAACproxydata %>% 
  dplyr::filter(Site %in% c("Cedar Island","Nassau",
                            "East River Marsh", "Swan Key",
                            "Placentia",
                            "Pelham Bay","Fox Hill Marsh",
                            "Snipe Key","Big River Marsh"))
\end{verbatim}

\normalsize

Next, the \texttt{reslr\_load} function is required for the preparation of input data for the NI GAM, which necessitates additional information not required by earlier models. Firstly, the statistical model relies on an estimate of the ``linear local rate'' and its associated uncertainty. By setting \texttt{include\_linear\_rate\ =\ TRUE}, the package incorporates this rate, which is assumed to stem from physical processes like Glacial Isostatic Adjustment (GIA). Users have the flexibility to include their preferred linear rate values as additional columns (\texttt{linear\_rate} and \texttt{linear\_rate\_err}) in the input dataframe. If these values are not provided, the package automatically calculates them using the available data.

Secondly, users are encouraged to include tide-gauge data by setting \texttt{include\_tide\_gauge\ =\ TRUE}. As discussed previously, users need to make a decision regarding the inclusion of the closest tide gauge (\texttt{TG\_minimum\_dist\_proxy\ =\ TRUE}), selecting specific tide gauges by providing a list of names (\texttt{list\_preferred\_TGs\ =\ c("ARGENTIA")}), or including all tide gauges within a one-degree proximity of the proxy site (\texttt{all\_TG\_1deg\ =\ TRUE}). Additionally, the tide-gauge data requires values for the \texttt{linear\_rate} and \texttt{linear\_rate\_err} columns, which are calculated using the ICE-5G (VM2) Model \citep{Peltier2004} with an uncertainty value of 0.3 mm/year \citep{Engelhart2009}, both provided within the \texttt{reslr} package.

Thirdly, the tide-gauge data is averaged over a decade to equation with the resolution of proxy records. If necessary, users can adjust the size of the averaging window to accommodate varying sediment accumulation rates. For example, a longer sediment accumulation rate would result in a larger average, such as 20 years. The default setting for \texttt{sediment\_average\_TG} is 10 years, which we will use in our example.

The final setting of the \texttt{reslr\_load} function is \texttt{prediction\_grid\_res}, allowing users to modify the resolution of the output plots. The default setting of 50 years serves as a starting point, but users have the flexibility to explore alternative options. For our example, we will utilize nine proxy sites and select all tide-gauges within a one-degree range of our proxy site, maximizing the number of data points to demonstrate the capabilities of our package. The specific settings employed are described below:
\footnotesize

\begin{verbatim}
multi_site_input <- reslr_load(
  data = multi_site,
  include_tide_gauge = TRUE,
  include_linear_rate = TRUE,
  TG_minimum_dist_proxy = TRUE,
  all_TG_1deg = TRUE)
\end{verbatim}

\normalsize

Similar to the previous example, the output of this function is a list of two dataframes called \texttt{data} and \texttt{data\_grid}. The \texttt{data} dataframe is the inputted data with additional columns for the data\_type\_id which will contain ``ProxyRecord'' and ``TideGaugeData''. The \texttt{data\_grid} is a dataframe that is evenly spaced in time based on the \texttt{prediction\_grid\_res} value chosen by the user and is used to create the plots. In this example, we have 9 proxy sites and 26 tide gauges and the data can be accessed by:
\footnotesize

\begin{verbatim}
glimpse(multi_site_input$data)
\end{verbatim}

\begin{verbatim}
#> Rows: 1,130
#> Columns: 14
#> $ Region          <chr> "Florida", "Florida", "Florida", "Florida", "Florida",~
#> $ Site            <chr> "Nassau", "Nassau", "Nassau", "Nassau", "Nassau", "Nas~
#> $ LongLat         <chr> "30.6_-81.7", "30.6_-81.7", "30.6_-81.7", "30.6_-81.7"~
#> $ Latitude        <dbl> 30.6, 30.6, 30.6, 30.6, 30.6, 30.6, 30.6, 30.6, 30.6, ~
#> $ Longitude       <dbl> -81.7, -81.7, -81.7, -81.7, -81.7, -81.7, -81.7, -81.7~
#> $ RSL             <dbl> 0.05, 0.03, 0.01, -0.01, -0.03, -0.05, -0.07, -0.09, -~
#> $ Age             <dbl> 2002, 1990, 1980, 1974, 1964, 1936, 1920, 1906, 1896, ~
#> $ Age_err         <dbl> 4.25, 5.50, 4.25, 4.50, 9.50, 10.75, 8.75, 9.75, 9.50,~
#> $ RSL_err         <dbl> 0.07, 0.07, 0.07, 0.07, 0.07, 0.07, 0.07, 0.07, 0.07, ~
#> $ SiteName        <fct> "Nassau,\n Florida", "Nassau,\n Florida", "Nassau,\n F~
#> $ data_type_id    <fct> ProxyRecord, ProxyRecord, ProxyRecord, ProxyRecord, Pr~
#> $ linear_rate     <dbl> 0.417923, 0.417923, 0.417923, 0.417923, 0.417923, 0.41~
#> $ linear_rate_err <dbl> 0.002958023, 0.002958023, 0.002958023, 0.002958023, 0.~
#> $ ICE5_GIA_slope  <dbl> 0.3241366, 0.3241366, 0.3241366, 0.3241366, 0.3241366,~
\end{verbatim}

\normalsize

A brief insight into the outputs of the \texttt{reslr\_input} function, e.g. number of observations and number of locations, can be obtained using the print function shown below:
\footnotesize

\begin{verbatim}
print(multi_site_input)
\end{verbatim}

\begin{verbatim}
#> This is a valid reslr input object with 1130 observations and  35 site(s).
#> There are  9  proxy site(s) and  26  tide gauge site(s).
#> The age units are; Common Era. 
#> Decadally averaged tide gauge data included by the package. 
#> The linear_rate and linear_rate_err has been included.
\end{verbatim}

\normalsize

A plot of the raw data can be created using \texttt{plot} function with an option to plot the tide gauges and the proxy records together or have present separate plots for each data source. Figure 4 demonstrates the resulting plot for the proxy records only using the following function:
\footnotesize

\begin{verbatim}
plot(x = multi_site_input,
     plot_proxy_records = TRUE,
     plot_tide_gauges = FALSE,
     plot_caption = FALSE)
\end{verbatim}

\begin{figure}[H]

{\centering \includegraphics[width=360px]{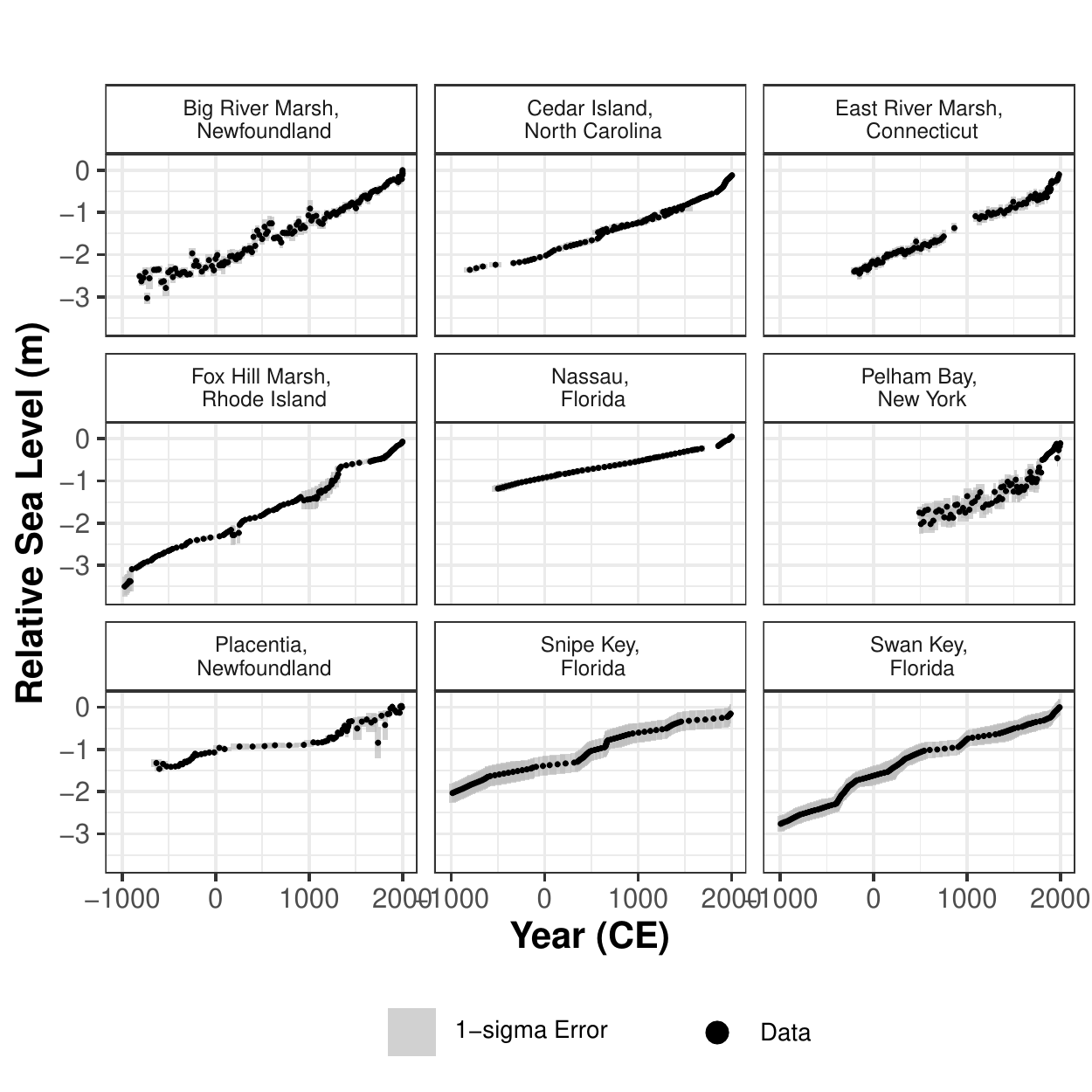} 

}

\caption{A plot of the raw data for our nine example sites along the Atlantic coast of North America. The x-axis is time in years in the Common Era (CE) and the y-axis is relative sea level in metres. The grey boxes are 1 standard deviation vertical and horizontal (temporal) uncertainty. The black dots are the midpoints of the uncertainty boxes.}\label{fig:plotdatamulti}
\end{figure}
\normalsize

For this model type, the \texttt{reslr\_mcmc} function should specify the \texttt{model\_type\ =\ "ni\_gam\_decomp"} and the MCMC simulation settings can be altered to ensure convergence.
\footnotesize

\begin{verbatim}
res_ni_gam_decomp <- reslr_mcmc(
  input_data =  multi_site_input,
  model_type = "ni_gam_decomp"
)
\end{verbatim}

\normalsize

The output of the \texttt{reslr\_mcmc} function is a list that stores the JAGS model run, the input dataframe and the dataframes for plotting the results. Identical to the other model processes, the convergence of the MCMC algorithm is examined and the parameter estimates from the model can be investigated using the following:
\footnotesize

\begin{verbatim}
summary(res_ni_gam_decomp)
\end{verbatim}

\begin{verbatim}
#> No convergence issues detected.
#> # A tibble: 4 x 7
#>   variable       mean      sd     mad     q5    q95  rhat
#>   <chr>         <num>   <num>   <num>  <num>  <num> <num>
#> 1 sigma_beta_h 1.82   0.257   0.247   1.45   2.29    1.00
#> 2 sigma_beta_r 0.288  0.0547  0.0505  0.213  0.391   1.00
#> 3 sigma_beta_l 0.939  0.148   0.145   0.717  1.21    1.00
#> 4 sigma_y      0.0155 0.00107 0.00108 0.0138 0.0173  1.00
\end{verbatim}

\normalsize

For the parameter estimates, we provide the standard deviation associated with each component of the NI GAM decomposition. Specifically, ``sigma\_beta\_r'' represents the standard deviation of the spline coefficient for the regional component, ``sigma\_beta\_l'' represents the standard deviation of the spline coefficient for the non-linear local component, ``sigma\_beta\_h'' denotes the standard deviation of the site-specific vertical offset component, and ``sigma\_y'' indicates the data model variation. These names correspond to the algebraic components described in Section 2 above.

One of the key advantages of the NI GAM approach is its ability to decompose regional RSL change into separate components. The results from the \texttt{ni\_gam\_decomp} model can be visualized using the \texttt{plot} function, which generates individual plots for each component. Additionally, all components, except for the linear local component, have corresponding rate plots. Users can access the data used to create each plot in the \texttt{res\_ni\_gam\_decomp} object as separate dataframes for each component.

In our example, we demonstrate the rate of change for the total model fit in Figure 5. This figure illustrates the rate of change at each site, which is useful to understand the variations of the relative sea-level signal, i.e. \(f(\mathbf{x},t)\). To plot the rate of change, users can employ the following method:
\footnotesize

\begin{verbatim}
plot(res_ni_gam_decomp, 
     plot_type = "rate_plot",
     plot_caption = FALSE)
\end{verbatim}

\begin{figure}[H]

{\centering \includegraphics[width=360px]{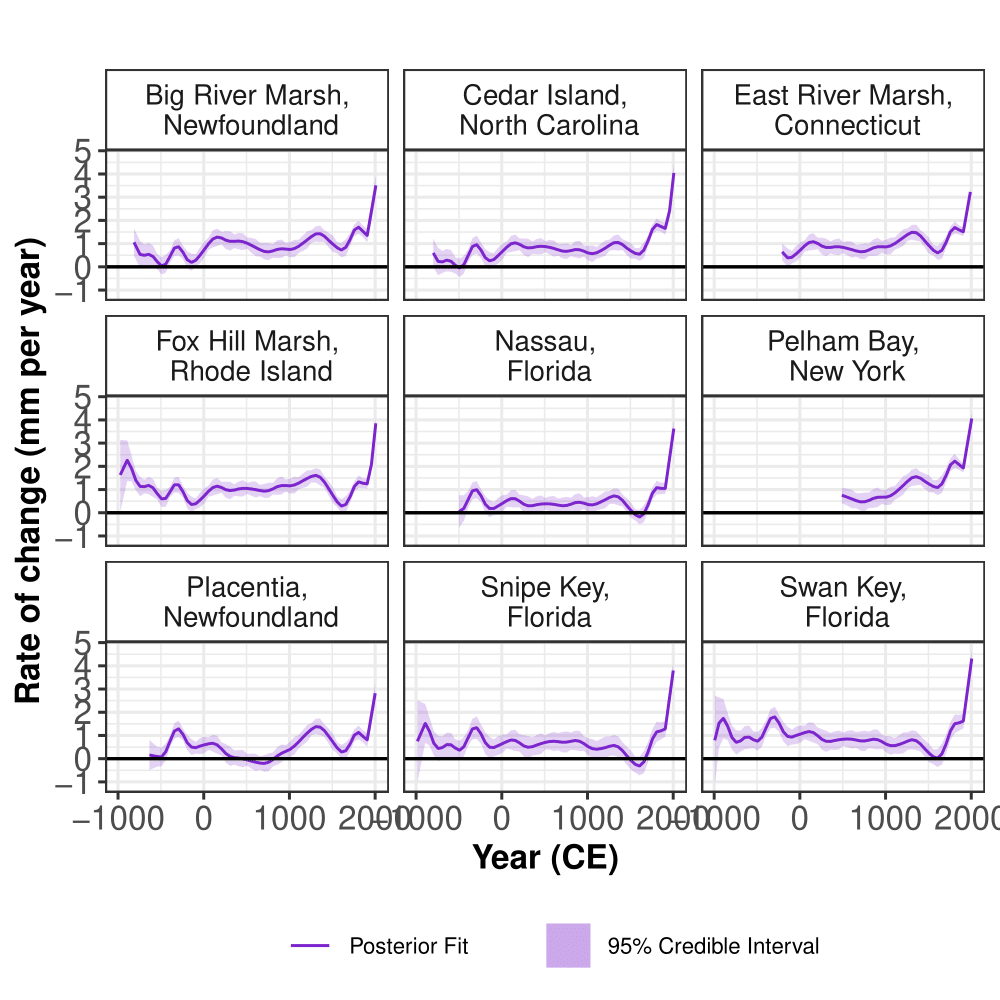} 

}

\caption{The rate of change for the total model fit for the noisy input generalised additive model for sites along the Atlantic coast of North America. It is calculated by finding the derivative of the total model fit. The solid purple line is the mean rate of change fit and the shading denotes 95\% credible interval for each site along the Atlantic coast of North America. The x-axis is time in years in the Common Era (CE) and the y-axis is rate of change in mm per year.}\label{fig:niplotrateload}
\end{figure}

\normalsize

The regional component (\(r(t)\)) captures the mean of RSL change along the Atlantic coast of North America. The associated rate of change of the regional component, as seen in Figure 6, provides an important visual insight into the rate at which this trend varied over the past 3000 years. It is accessed by:
\footnotesize

\begin{verbatim}
plot(res_ni_gam_decomp, 
     plot_type = "regional_rate_plot",
     plot_caption = FALSE)
\end{verbatim}

\begin{figure}[H]
{\centering \includegraphics[width=360px]{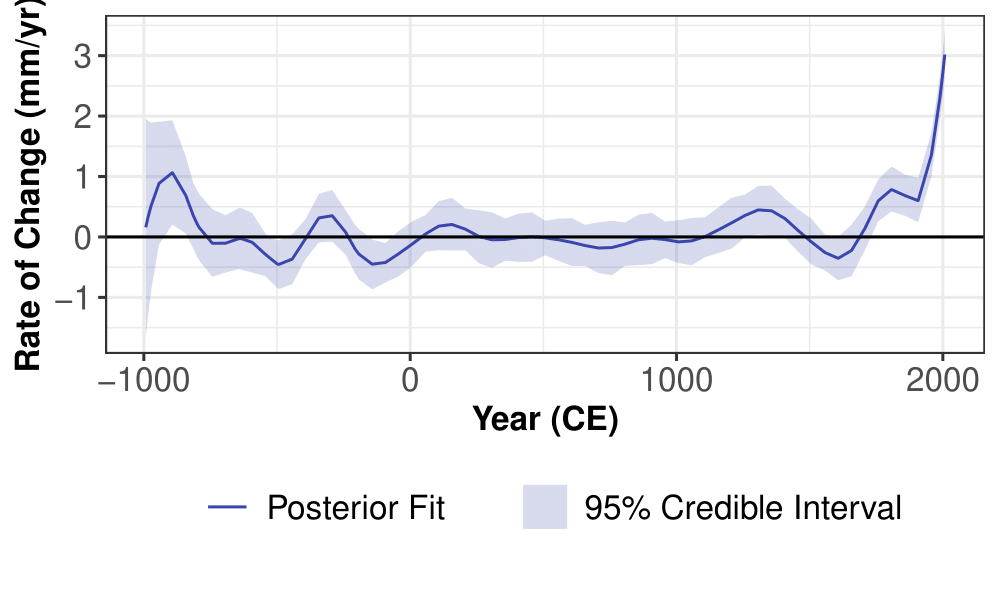} 
}
\caption{The rate of change for the regional component of the noisy input generalised additive model for the nine proxy sites and the eleven tide gauges along the Atlantic coast of North America. It is calculated by finding the derivative of the regional component fit. The solid blue line is the mean rate of change fit and the shading denotes 95\% credible interval. The x-axis is time in years in the Common Era (CE) and the y-axis is rate of change in mm per year.}\label{fig:regrateplotload}
\end{figure}

\normalsize

\hypertarget{summmary}{%
\section{Summary}\label{summmary}}

In this paper, we have presented an overview of the \texttt{reslr} package and discussed its various features and design decisions. Our goal was to address the specific needs of the paleo sea-level community and provide an efficient and flexible R package that caters to different types of source data, whilst maintaining a simple workflow that does not require the user to learn too many different functions.

Through two case studies, we demonstrated the simplicity and accessibility of the package. The first case study examined a single site using the NI spline in time. Our results showed that the \texttt{reslr} package can provide RSL estimates and associated rate of change values over time for a single location. In the second case study, we showcased the capabilities of the \texttt{reslr} package when analysing data from multiple locations. We highlighted its flexibility, allowing for the decomposition of the relative sea-level signal into different components. Additionally, we presented a comprehensive method for incorporating tide-gauge data, which can help to provide valuable insights into recent changes in RSL not captured by proxy records.

There are several potential extensions for the \texttt{reslr} package. One possibility is to include additional statistical models, such as machine learning techniques, to accommodate larger datasets. Another improvement could be the integration of other instrumental data sources, such as satellite data, enabling the examination of other variables related to climate change. Overall, the \texttt{reslr} package offers a powerful toolkit for the paleo sea-level community, and we anticipate that it will continue to evolve and expand its capability to meet the evolving needs of researchers in this field.

\hypertarget{acknowledgements}{%
\section{Acknowledgements}\label{acknowledgements}}

Upton's work is supported by A4 (Aigéin, Aeráid, agus athrú Atlantaigh) project is funded by the Marine Institute (grant: PBA/CC/18/01). Parnell's work is supported by the SFI awards 17/CDA/4695; 16/IA/4520; 12/RC/2289P2. Cahill's research is conducted with the financial support of Science Foundation Ireland and co-funded by Geological Survey Ireland under Grant number 20/FFP-P/8610.
\newpage
\hypertarget{appendix}{%
\section{Appendix}\label{appendix}}

\hypertarget{exampledatasetappendix}{%
\subsection{Example dataset}\label{exampledatasetappendix}}

The \texttt{reslr} package contains a dataset used as an example called \texttt{NAACproxydata}. This dataset contains proxy records from the Atlantic coast of North America as used in \citet{Upton2023noisy}. The 21 different proxy data sites and the references for each data source can be found in Table 2. 

\begin{table}[H]
\centering
\begin{tabular}{p{50mm}l}
  \hline
  Site Name & Reference\\
  \hline
Barn Island, Connecticut & \citet{Donnelly2004,Gehrels2020}\\
Big River Marsh, Newfoundland & \citet{Kemp2018} \\
Cape May Courthouse, New Jersey & \citet{Kemp2013SealevelUSA,Cahill2016}\\
Cedar Island, North Carolina & \cite{Kemp2011,Kemp2017} \\
Cheesequake, New Jersey & \citet{Walker2021} \\
Chezzetcook Inlet, Nova Scotia & \citet{Gehrels2020} \\
East River Marsh, Connecticut & \citet{Kemp2015,Stearns2023}\\
Fox Hill Marsh, Rhode Island & \citet{Stearns2023} \\
Leeds Point, New Jersey & \citet{Kemp2013SealevelUSA,Cahill2016} \\
Les Sillons, Magdelen Islands & \citet{Barnett2017} \\
Little Manatee River, Florida & \citet{Gerlach2017} \\
Nassau, Florida & \citet{Kemp2014} \\
Pelham Bay, New York & \citet{Kemp2017,Stearns2017} \\
Placentia, Newfoundland & \citet{Kemp2018} \\
Revere, Massachusetts & \citet{Donnelly2006} \\
Saint Simeon, Quebec & \citet{Barnett2017} \\
Sanborn Cove, Maine & \citet{Gehrels2020} \\
Sand Point, North Carolina & \citet{Kemp2011,Kemp2017} \\
Snipe Key, Florida & \citet{Khan2022} \\
Swan Key, Florida & \citet{Khan2022} \\
Wood Island, Massachusetts & \citet{Kemp2011}\\
 \hline
\end{tabular}
\caption{Presents the names of all the sites available in the example dataset within the \texttt{reslr} package. For each site we include the reference in the literature to the source of the data. \label{Tab:data_ref}}
\end{table}

\addcontentsline{toc}{section}{References}
\renewcommand\bibname{Reference}
\bibliography{RJreferences.bib}

\address{%
Maeve Upton\\
Maynooth University\\%
Hamilton Insititute, ICARUS and Department of Mathematics and Statistics\\ Maynooth, Ireland\\
\textit{ORCiD: \href{https://orcid.org/0000-0002-1151-6657}{0000-0002-1151-6657}}\\%
\href{mailto:uptonmaeve010@gmail.com}{\nolinkurl{uptonmaeve010@gmail.com}}%
}

\address{%
Andrew Parnell\\
Maynooth University\\%
Hamilton Institute, ICARUS and Department of Mathematics and Statistics\\ Maynooth, Ireland\\
\textit{ORCiD: \href{https://orcid.org/0000-0001-7956-7939}{0000-0001-7956-7939}}\\%
\href{mailto:andrew.parnell@mu.ie}{\nolinkurl{andrew.parnell@mu.ie}}%
}

\address{%
Niamh Cahill\\
Maynooth University\\%
Department of Mathematics and Statistics and ICARUS\\ Maynooth, Ireland\\
\textit{ORCiD: \href{https://orcid.org/0000-0003-3086-550X}{0000-0003-3086-550X}}\\%
\href{mailto:niamh.cahill@mu.ie}{\nolinkurl{niamh.cahill@mu.ie}}%
}

\end{article}

\end{document}